\documentclass[conference]{IEEEtran}

\ifCLASSINFOpdf
\else
\fi

\usepackage{amsmath}
\usepackage{cite}
\usepackage{graphicx}
\usepackage{tabularx}
\usepackage{array}
\usepackage[linesnumbered,ruled,vlined]{algorithm2e}

\usepackage{subcaption}
\usepackage{booktabs}

\DeclareUnicodeCharacter{2212}{-}

\begin{document}
%
\title{Dynamic Digital Twins of Blockchain Systems: State Extraction and Mirroring}


%
\author{
    \IEEEauthorblockN{
        Georgios Diamantopoulos\IEEEauthorrefmark{1}\IEEEauthorrefmark{2},
        Nikos Tziritas\IEEEauthorrefmark{3},
        Rami Bahsoon\IEEEauthorrefmark{2}, 
        Nan Zhang\IEEEauthorrefmark{1}, 
        Georgios Theodoropoulos\IEEEauthorrefmark{4}\IEEEauthorrefmark{1}
    }
\\
\IEEEauthorblockA{
    \IEEEauthorrefmark{1}
        Southern University of Science and Technology (SUSTech), Shenzhen, China}
\IEEEauthorblockA{
    \IEEEauthorrefmark{2}
        University of Birmingham, Birmingham, United Kingdom}
\IEEEauthorblockA{
    \IEEEauthorrefmark{3}
        University of Thessaly, Lamia, Greece}
\IEEEauthorblockA{
    \IEEEauthorrefmark{4}
        Research Institute of Trustworthy Autonomous Systems, SUSTech, Shenzhen, China}
}


\maketitle

\begin{abstract}
Blockchain adoption is reaching an all-time high, with a plethora of blockchain architectures being developed to cover the needs of applications eager to integrate blockchain into their operations. However, blockchain systems suffer from the trilemma trade-off problem, which limits their ability to scale without sacrificing essential metrics such as decentralisation and security. The balance of the trilemma trade-off is primarily dictated by the consensus protocol used. Since consensus protocols are designed to function well under specific system conditions, and consequently, due to the blockchain's complex and dynamic nature, systems operating under a single consensus protocol are bound to face periods of inefficiency. The work presented in this paper constitutes part of an effort to design a Digital Twin-based blockchain management framework to balance the trilemma trade-off problem, which aims to adapt the consensus process to fit the conditions of the underlying system. Specifically, this work addresses the problems of extracting the blockchain system and mirroring it in its digital twin by proposing algorithms that overcome the challenges posed by blockchains' decentralised and asynchronous nature and the fundamental problems of global state and synchronisation in such systems. The robustness of the proposed algorithms is experimentally evaluated.
\end{abstract}

\begin{IEEEkeywords}
Blockchain, Digital Twin, DDDAS, State Replication, Synchronisation
\end{IEEEkeywords}

%
\IEEEpeerreviewmaketitle

\section{Introduction}
\label{Introduction}

In the last decade, blockchain technology has seen a tremendous increase in popularity. Initially proposed as the core of the now well-known decentralised currency platform Bitcoin \cite{btc} in 2008, blockchain has since evolved and grown in adoption rapidly. Despite blockchain's most broadly known applications being in finance, a version designed for smaller-scale permissioned networks, thus named \textit{permissioned blockchain}, is emerging. Permissioned blockchain has been used in Supply Chains, Smart Grids, e-government and many more fields \cite{gad2022emerging} for its decentralisation, security and immutability properties.

Intuitively, one can think of blockchain as a replicated database where each user (node) holds a complete trace of the system state changes (transactions) organised in a time-ordered linked list of blocks, hence the name `blockchain'. To achieve the above, nodes communicate over a peer-to-peer (p2p) network following a byzantine fault-tolerant state machine replication algorithm (consensus protocol). Blockchain can offer unmatched decentralisation, transparency and security by replicating the entire state transition history amongst all nodes using a byzantine fault-tolerant protocol and cryptography. Unfortunately, the above comes at the cost of performance, as the mechanisms required to achieve these properties increase the system's complexity.

The problem of balancing scalability, security, and decentralisation is referred to as the \textit{trilemma trade-off problem} of blockchain, a term coined by Ethereum's Vitalik Buterin; it implies that in a blockchain system, the above three properties are inversely correlated. The consensus protocol plays a central role in shaping these properties, as it is the component that controls the process of adding new blocks, i.e., updating the database, in a decentralised way. Besides the design of the consensus protocol, the underlying P2P network and the complexity of the application can also heavily affect its performance and, thus, that of the entire blockchain system.

A critical step in the design phase of any blockchain system is choosing an appropriate consensus protocol, considering the environment and workload that will be placed upon the system, to ensure maximum scalability while fulfilling the security and decentralisation requirements. Unfortunately, blockchains are complex and dynamic in nature, and consensus protocols are highly specialised algorithms designed to function well under specific system conditions. Consequently, a blockchain system utilising a single consensus protocol is bound to face periods of sub-optimal performance or, worse, periods vulnerable to attacks.

Therefore, a dynamic reconfiguration approach can help balance the trilemma trade-off problem by ensuring that a consensus protocol suitable to the underlying system's state is always used. 
To address this challenge in previous work, we proposed a Digital Twin-based approach for the dynamic management of blockchain systems to optimise their trilemma trade-off problem. In \cite{10015447,10305798}, we presented the architecture of our proposed Digital Twin, outlining its basic constituent components, and in \cite{10.1007/978-3-031-52670-1_28}, we demonstrated the utilisation of the Digital Twin for the dynamic selection of consensus protocols to optimise the scalability of the system through the incorporation of a Reinforcement Learning agent. At the heart of the Digital Twin is a typical DDDAS feedback loop between the blockchain system and the Digital Twin. In  \cite{10.1145/3573900.3591121}, we presented SymBChainSim, a novel simulation system for blockchain systems in the context of the proposed Digital Twin architecture. SymBChainSim is DDDAS-compliant as it allows and supports the update of the simulation model of the blockchain in real-time based on information extracted from the physical system. As a continuation of this work, this paper aims to address the following two interrelated challenges:
  
\begin{enumerate}
    \item  How can the Digital Twin extract state information from the blockchain system, given the latter's decentralised, asynchronous nature? 
    \item How can the Digital Twin use this information to mirror the system's state and maintain an up-to-date model of the system?   
\end{enumerate}


The rest of the paper is structured as follows: Section \ref{System Model} presents a model of a blockchain system and the metrics relevant to the trilemma trade-off. Section \ref{Twinning a Blockchain} discusses the challenges of twining a blockchain system and presents conceptual methods for blockchain state extraction. Section \ref{Peer State Estimation} presents the proposed state extraction method and addresses issues caused by the asynchronous nature of blockchain. Finally, section \ref{Experimental Evaluation} presents the experimental evaluation of the proposed method with section \ref{Peer State Estimation} concluding this paper.




\section{System Model}
\label{System Model}

\begin{figure*}[!h]
  \centering
  \includegraphics[width=.8\linewidth]{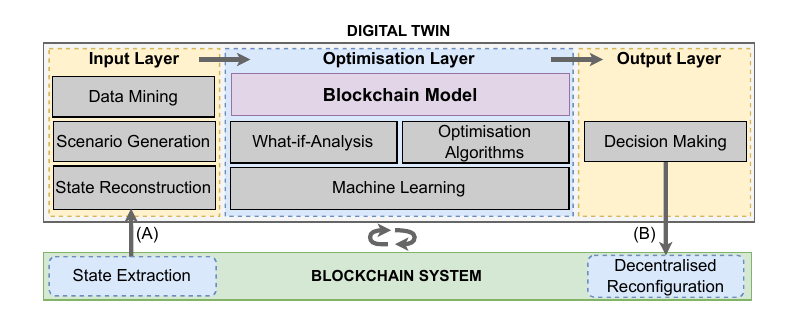}
  \caption{High-level architecture of proposed digital twin-based blockchain management framework.}
  \label{fig:DT_arch}
\end{figure*}





\subsection{Blockchain}


Let $S$ denote a blockchain system as a set of nodes with the $i-th$ node of $S$ denoted as $n_i$. Each node $n_i$ in $S$ is connected with $P_{i}$ other nodes over bi-directional network links $L$ with $l_{i,j}$ denoting the link between the nodes $n_i$ and $n_j$. Each network link $l$ is defined by a latency value denoted $D(l)$ and a bandwidth value $B(l)$. The nodes in $P_i$ are called peer nodes of $n_i$ and $P_i \subset S, \forall i \in S$. Peer nodes can use their link to exchange messages $m_{i,j}$ with $i$ denoting the sender and $j$ the receiver. The size of a message is denoted as $S(m)$. Since $P_i \subset S, \forall I \in S$, messages are propagated using gossip (also known as epidemic) protocols utilising cascading peer-multi-casts. This process is modelled after the spread of epidemics and ensures that messages eventually reach all available nodes.

Blockchain systems are essentially databases that store data from an underlying application. The database entries are in the form of transactions which describe database operations. Let $t_i$ denote the $i-th$ transaction arriving at the blockchain system from the underlying application, with $T(t)$ denoting its arrival and $S(t)$ its size. To allow for decentralised state updates, a consensus protocol $CP$ is used for the nodes to agree on the order and validity of transactions. Rather than reaching a consensus on individual transactions nodes take turns creating and proposing blocks of transactions. Each block in the blockchain system references its preceding block, creating a back-linked list of blocks storing the entire database operation history. 

More formally, let $C$ denote the back-linked list of blocks accepted and stored locally by all $n \in S$ with $b_i$ denoting the $i-th$ block. $T(b_i)$ denotes the time $B_i$ was added to $C$. $Tx(B_i)$ denotes the set of all transactions included in $b_i$ and $S(b_i)$ the size of $b_i$ and $M(b_i)$ denotes the node (miner/proposer) that created $b_i$.

In permissioned blockchain systems, only a special set of nodes in $S$ called block producers can participate in $CP$ to create new blocks. A $CP$ defines a communication pattern the nodes must follow and respond to according to their local state. This communication pattern occurs in rounds and is initialised by a node proposing a block of transactions to the system. If the criteria specified by the $CP$ are met by the end of a round, the block is accepted, and nodes must add it to their local block list $C$ (blockchain).

\subsection{Trilemma Trade-off Properties}

\textbf{Scalability} refers to the number of nodes and transactions a blockchain system can support while maintaining good Quality of Experience (QoE). Notable KPIs for measuring scalability are transaction latency and throughput, with the former measuring the transaction processing time and the latter the number of transactions the blockchain system can process per second. While transaction latency is the most important factor affecting the QoE, as it directly measures the system's response time, it partially depends on the system throughput. If the number of transactions arriving in the system exceeds the throughput, transaction latency will gradually decline as the backlog of unprocessed transactions builds up.

\textbf{Decentralisation} refers to the distribution of blocks over the block-producing nodes of the system. A uniform distribution means block producers have an equal chance to produce a block and thus have an equal say over the evolution of the system's state. Decentralisation is a core property of blockchain systems; therefore, it is critical to maintain it at high levels. The Gini coefficient, proposed initially to measure income inequality, can also be used to calculate the decentralisation levels of a blockchain system. Specifically, the authors of \cite{liu2019performance} propose a modified Gini coefficient to measure the inequality of block production in a blockchain system. 

\textbf{Security} refers to the resilience of a blockchain system against attacks. Although security covers a vast spectrum, attacks against the consensus process are considered in the context of the trilemma. These attacks are initiated by malicious nodes in the network and can target the system's decentralisation and scalability. Attacks on decentralisation aim to give the malicious node/s control of the system, while attacks on scalability aim to reduce the blockchain system's QoE drastically.

The above properties are expressed by the characteristics of the consensus process, which are, in turn,  defined by the consensus protocol, the network state, the system workload, and the behaviour of the system nodes. Consequently, a digital twin with an accurate model of the consensus process can act as a platform for optimising the trilemma trade-off problem of blockchain.

\section{Twinning a Blockchain}
\label{Twinning a Blockchain}

Digital twins are essentially Dynamic Data-Driven Application Systems (DDDAS) that form a real-time info-symbiotic feedback loop with a physical system to create and maintain a digital model of it. Digital twins offer insight into the inner workings of the physical system and provide a platform through which optimisation decisions and hypothetical scenarios can be evaluated and applied to the physical system.



In previous work, we have explored the use of digital twins to balance the trilemma trade-off problem of permissioned blockchain systems \cite{10015447, 10.1007/978-3-031-52670-1_28}. A high-level view of architecture for digital twin-based permissioned blockchain management was proposed in \cite{10015447} and can be seen in Fig. \ref{fig:DT_arch}. Besides the challenges associated with the decision-making process, creating a digital twin for optimising a blockchain system requires the construction of a bidirectional feedback loop. Specifically, a feedback loop requires  (A) extracting and constructing the state of the physical system and (B) feeding the optimisation decisions back to the physical system and updating its configuration accordingly (Fig. \ref{fig:DT_arch}); This paper focuses on the former problem (A). 

The prerequisite of valid analysis enabled by digital twins is maintaining an equivalent physical system model. The system's state (current status or condition) must be updated periodically with parameters defining the dynamics of state evolution over time, requiring continuous calibration. The frequency of the updates plays a crucial role in both the model's accuracy and, conversely, the communication and computation overhead involved \cite{zhang2024knowledge,tan_digital_2023}. In \cite{zhang2024knowledge}, we examined and addressed this challenge by proposing novel approaches for adaptive and cost-efficient model updates for intelligent systems. 

\subsection{Extracting the Blockchain State}
\label{sec:Extract}

\begin{figure*}
    \centering
    \includegraphics[width=1\linewidth]{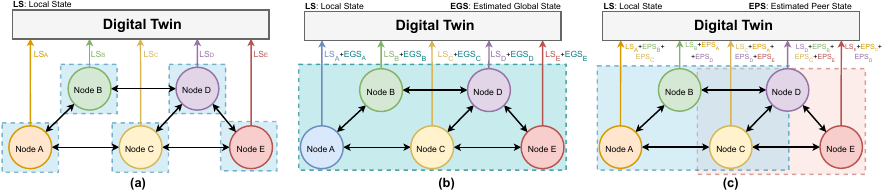}
    \caption{Blockchain state extraction approaches.}
     \label{fig:graphs}
\end{figure*}

Blockchain poses unique challenges that must be overcome to construct a DDDAS-type feedback loop between a digital twin and a blockchain system. The combination of its decentralised nature, asynchronous network, and the possible presence of byzantine nodes makes extracting the state of a blockchain system far from trivial. Since blockchain networks are decentralised, no central authority exists to provide the twin with a definitive `global' blockchain state. Instead, the `global' blockchain state exists only as the combination of the `local' states of the blockchain system nodes.

\subsection{Related Work:}
Blockchain-enabled monitoring has been utilised as an approach to store the monitored state of blockchain-supported applications systems \cite{choi2020novel, junior2022blockchain, alam2023overview} including Digital Twins  \cite{hasan2020blockchain, mandolla2019building, putz2021ethertwin};  this latter case refers to blockchain for Digital Twins rather than Digital Twins of blockchain systems, which is the subject of this paper. 


 More related to the work presented in this paper are systems focusing on monitoring the blockchain system itself. Such systems focus on the problem of blockchain performance monitoring. In \cite{9793843}  BCmaster, a blockchain performance monitoring framework utilising Prometheus and Grafana to extract and visualise performance metrics for IoT Blockchains is presented. Specifically, the performance metrics are collected by requiring the execution of the blockchain node to take place in a docker container monitored by BCmaster's data parser. The data parsers submit their data to a central visualiser through Prometheus, which utilises the Grafana dashboard to visualise the system's performance. In \cite{10.1145/3183519.3183546}, the authors propose a mechanism for extracting performance information from blockchain nodes by analysing the logs produced during execution. Similar to \cite{9793843}, the logs are parsed by daemon services running locally on the systems executing the node logic. The results are sent to a centralised service for visualisation.  
 
 Although these approaches provide mechanisms for blockchain monitoring, two main limitations limit their use in a Digital Twin context, such as the one presented in this paper. Firstly, they do not address the real-time performance requirements of Digital Twin systems. More importantly, they do not provide any mechanisms to guarantee reliable and accurate state mirroring, which is essential in the context of Digital Twins. In the following sections, we discuss these issues and propose novel solutions.   




\subsection{Local State Extraction}

A straightforward approach to extracting the state of a blockchain system is one in which each node sends its local state to the digital twin, such as the one depicted in Fig. \ref{fig:graphs}a. This approach works well when the network behaves synchronously; thus, data transmission is timely, and messages are not lost. Furthermore, this approach is efficient as it requires minimal data transmission between the nodes. Unfortunately, this approach breaks down at the first sign of network asynchrony, as any lost or delayed messages make reconstructing the global blockchain state impossible. 

\subsection{Global State Estimation}

While an extraction mechanism relying on nodes reporting their local state cannot tolerate the asynchronous nature of blockchain networks, the underlying mechanisms of blockchain may be utilised to allow for this. The above is enabled because blockchain nodes participate in a strictly defined consensus process through which frequent communication occurs. Consequently, a single blockchain node can attempt to estimate a global blockchain state by utilising information extracted by analysing network traffic and message contents. This approach is depicted in Fig. \ref{fig:graphs}b.

By utilising the estimated global states received by the nodes, even under an asynchronous network, lost or delayed information can be reconstructed from the global state estimation contained in the messages that reached the twin in a timely manner. One important aspect to consider is that blockchain networks are not fully connected. Since blockchains are implemented over peer-to-peer networks, the system nodes are connected to a subset of peers. As a result, extracting the state of non-peer nodes, especially ones many network hops away, is challenging.

\subsection{Peer State Estimation}

Although estimating the global blockchain system is challenging due to the nature of blockchain networks, estimating the state of a node's immediate peers is possible due to their frequent communication. Based on this, a version of the mechanism shown in Fig.\ref{fig:graphs}b can be constructed in which the nodes transmit information about the state of their peers rather than the state of the entire blockchain network. This approach is depicted in Fig. \ref{fig:graphs}c. The above allows for state reconstruction under an asynchronous network but requires the state message of at least one immediate peer to reach the digital twin. In the rest of the paper, we focus on this approach.
\section{Peer State Estimation}
\label{Peer State Estimation}

\begin{figure*}
  \centering
  \includegraphics[width=.8\linewidth]{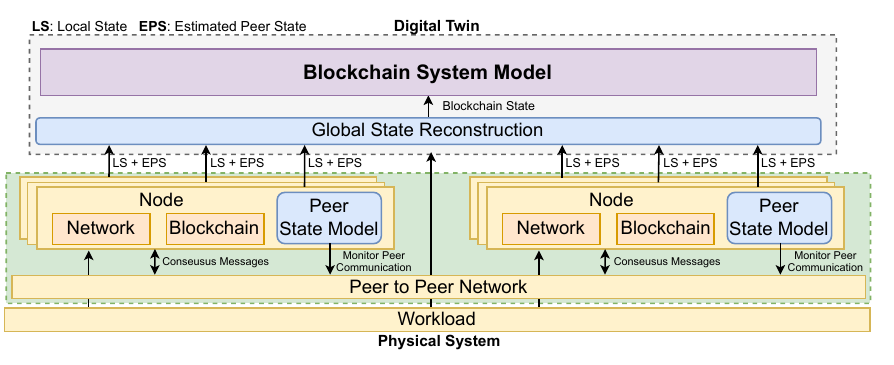}
  \vspace{-.2cm}
  \caption{Architecture of proposed blockchain state extraction mechanism based on estimated peer states}
  \label{fig:figure_label}
\end{figure*}

This section presents a mechanism for blockchain state extraction based on the estimated peer state model described in the previous section. The proposed mechanism relies on analysing the communication patterns and network traffic between peer nodes to estimate the peer's network state. This information can, thus, be utilised by the digital twin to reconstruct the global blockchain state in the event of lost or delayed messages. Specifically, the blockchain network's topology and each node's network state are required to simulate the consensus processes. Using the above, the propagation of messages through the network can be modelled and used to simulate the execution of the consensus protocol for the production of blocks. 

The local state of a node $n_i$ consists of its network bandwidth $BW(n_i)$ and a list of peers with which an active connection is maintained $P(n_i)$. The bandwidth of a network link between two peers is equal to that of the peer with the lowest bandwidth.
The estimated peer state of a node consists of the estimated bandwidth of all its peers
\begin{equation*}
        EPS(n_i) = \{EBW(n_i, n), \forall n \in P(n_i)\}
\end{equation*}
where $EBW(n_i, n)$ is the estimated bandwidth of $n$ by $n_i$ obtained by analysing their exchanged consensus messages. Specifically, $EBW(n_i, n_j)$ denotes the average bandwidth observed from message exchanges between $n_i$ and $n_j$ and is defined as 
\begin{equation}
        EBW(n_i, n_j) = \frac{\sum_{m \in M_{i,j}}\frac{S(m)}{D(m)}}{|M_{i,j}|}
        \label{eq:const_eps}
\end{equation}
where $M_{i,j}$ is the set of messages between $n_i$ and $n_j$ and $S(m)$ the size and $D(m)$ the transmission delay of $m$.

\begin{algorithm}[t]
    \caption{Global state reconstruction}
    \label{alg:gsr}
    \KwIn{received state messages $SM$}
    \KwOut{reconstructed blockchain state $S$}
    \Begin{
        Build the list of nodes $N$ using the local states $\in SM$ and their reported peers \\
        \For{$n \in N$}{ 
            \If{message from $n$ in $SM$}{
                $S(n) \gets$ local state in $SM_n$ \\
            }
            \Else{
                 $S_{peers}(n) \gets$ nodes with $n$ as peer in $M$ \\
                 $ES_n \gets$ all $EBW(n)$ from $EPS$ in $M$ \\
                 $S_{network}(n) \gets$ MAX($ES_n$) \\
            }    
        }
    }
\end{algorithm}

Using the above information, an approximation of the blockchain state can be constructed even in cases where the local states of the nodes are delayed or lost due to the asynchronous network. In cases where $N$ states are available when the blockchain model requires an update, the blockchain state can be constructed utilising the local state reported by the individual nodes. In cases where only $N-x$ states are available, the missing $x$ states can be attempted to be reconstructed by utilising the $N-x$, $EPS$ values received. Let $SM$ be the state messages received by the digital twin with $sm_i = (BW(n_i), P(n_i), EPS(n_i))$. The number of missing messages $x$ can be defined by observing the number of unique nodes that appear in all lists of peers in $M$ and comparing it to the number of state messages received $|SM|$. If the digital twin has received fewer messages than the number of unique nodes identified, algorithm \ref{alg:gsr} can be used to reconstruct their state. Specifically, the maximum estimated bandwidth is used as the link speed between two peers is equal to the lowest bandwidth between the two. Consequently, the bandwidth of a node acts as an upper bound to its bandwidth estimations; thus, the maximum estimated bandwidth for a node is the one that most closely reflects reality. 


\subsection{Synchronisation Issues}
\label{Synchronisation Issues}
\begin{figure*}
    \begin{subfigure}[b]{.45\linewidth}
        \centering
        \includegraphics[width=.8\textwidth]{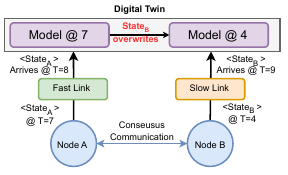} 
        \caption{No synchronisation}
        \label{sub:no_sync}
    \end{subfigure}
    \hfill
    \begin{subfigure}[b]{.45\linewidth}
        \centering
        \includegraphics[width=.8\textwidth]{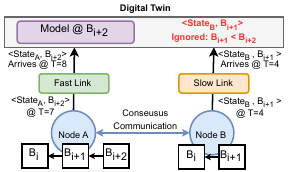} 
        \caption{Block Synchronisation}
        \label{sub:sync}
    \end{subfigure}
    \caption{Reconstructing a blockchain state with (b) and without (a) synchronisation.}
\end{figure*}


The previous section discussed the problem of reconstructing and mirroring the state of the blockchain system in the Digital Twin when local state updates from the blockchain system are missing.  The asynchronous nature of the system, however, poses an additional challenge. The order of state updates the Digital Twin receives is not necessarily the same as the order in which these messages are generated and sent by the blockchain neighbour nodes. The Digital Twin needs to ensure that it considers only the most recent state updates, or else the model may be updated with outdated information. 

This is illustrated in the example of Fig. \ref{sub:no_sync}. In this example, Node A sends its (peer) state information message $State_A$ at T=7, after its neighbour Node B sends its $State_B$ at T=4. In reality, $State_A$ represents the most recent state of the blockchain system; however, due to network delays, the Digital Twin receives $State_A$ first, updates the model and then receives $State_B$ and wrongly updates the state to reflect a past state. 

Synchronising local clocks and constructing global clocks in distributed systems is a challenging albeit well-studied problem \cite{10.5555/562145}. In this paper, we propose a solution that avoids the overheads involved in such existing approaches by utilising the blockchain structure to establish a common temporal framework for the blockchain nodes. 

\begin{algorithm}[t]
    \caption{Block-order synchronisation}
    \label{alg:sync}
   
        \KwIn{received state message $S$,}
        \KwOut{updated model $M$ based on $S_r$}
        \Begin{
            $B_{M} \gets$ block referenced by $S$\\
            \For{$n \in S$}{ 
                $B_{n} \gets$ block referenced in latest update of $n$\\
                \If{$B_{n} < B_{S}$}{
                    \tcc{\scriptsize $S$ contains a more recent state of $n$}
                    $M_n \gets S_n$ (based on Alg. \ref{alg:gsr})\\
                } \Else{
                    \tcc{\scriptsize $S$ contains an older state of $n$}
                    ignore $S_n$ \\ 
                }
                
            }
        }   
\end{algorithm}

In a blockchain system, the nodes participate in a consensus process through which new blocks to be stored in an immutable chain-like data structure are agreed upon. These blocks are constantly produced and propagated over the network with a $IBT$ (Inter-Block Time) frequency defined by the efficiency of the consensus process. Since the blockchain is immutable and new blocks are added in a monotonically increasing order for all nodes, the latest block in the blockchain of nodes can be used by digital twins to order state messages.

The above can be achieved by requiring state messages sent to the digital twin to include a reference to the latest block of the sender's blockchain (block header). By ordering state messages according to the block generation order in the blockchain, the digital twin can ensure that updates always utilise the most recent state information. It is important to consider that, unlike time, blocks are discrete, and thus, there exists a time window of uncertainty in which messages with the same block reference can arrive. Fortunately, this time window is bounded by $IBT$, which is typically small (in the range of a few seconds). Based on this, state messages referencing the same block as the most recent state update are ignored to reduce update overheads in the digital twin.
From the above, an interesting property emerges in digital twins that aim to optimise a blockchain system's consensus process. Since the efficiency of the consensus process affects the $IBT$, better decisions lead to a lower $IBT$. Consequently, such decisions improve the model's accuracy by reducing the uncertainty time window ($IBT$). Conversely, the opposite is also true. The proposed synchronisation approach is illustrated in algorithm \ref{alg:sync} and is demonstrated by the example in Fig. \ref{sub:no_sync}.
   
\subsection{Effect of Byzantine Behaviour}

While the above mechanisms can allow the digital twin to reconstruct an approximation of the global blockchain state in cases where messages are lost due to poor network conditions or offline nodes, another element can disrupt the state extraction and reconstruction process: the presence of malicious nodes that intentionally provide false information about their local state. Although malicious nodes are expected more in permissionless blockchains, such behaviour can potentially also occur in permissioned systems. This is an interesting problem, however, it is considered outside the scope of this paper; work in this area is already ongoing by the authors and will be presented in a future publication.


\section{Experimental Evaluation}
\label{Experimental Evaluation}

\begin{figure*}
    \begin{minipage}[c]{\textwidth}
        \begin{subfigure}[b]{\textwidth}
            \centering
            \includegraphics[width=.85\textwidth]{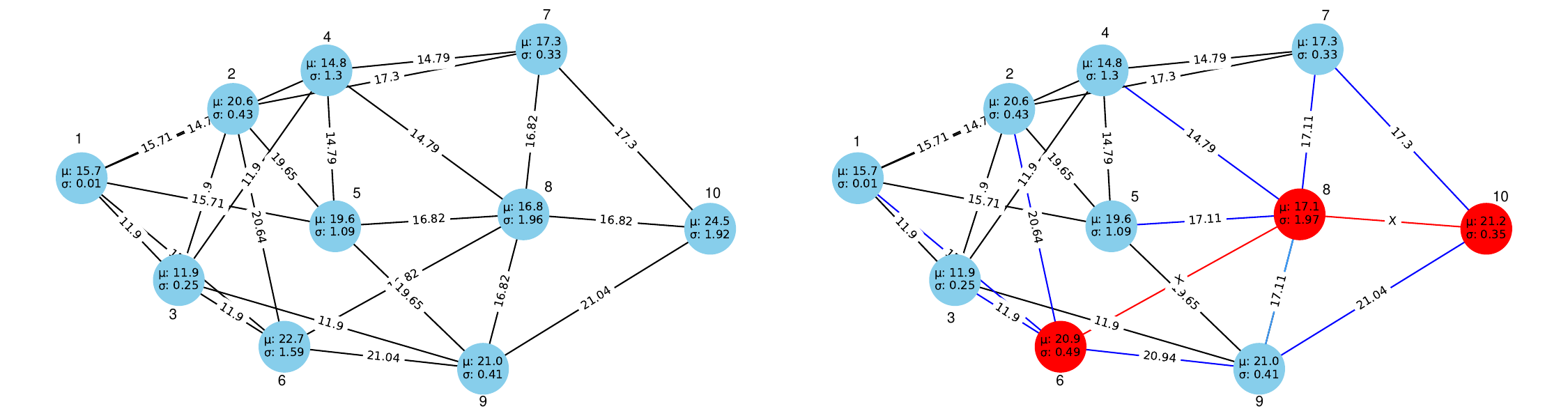}
            \caption{Visualisation of state reconstruction. Original (left) VS reconstructed (right). Red nodes failed to provide state information and are reconstructed using the estimated peer state of their peers. Blue edges represent reconstructed edges. Red edges could not be reconstructed. Node labels denote the local network state as a bandwidth distribution. Edge labels denote the bandwidth of the link.}
            \label{sub:Viz}
        \end{subfigure}
    \end{minipage}
    \begin{minipage}[c]{1\textwidth}
        \vspace{.1cm}
        \centering
        \resizebox{.55\textwidth}{!}{%
            \begin{tabular}{|l|c|c|c|c|c|c|c|c|}
                \toprule
                \textbf{Param} & $CP$ & Nodes & Peers per node & Num. Tx. & Tx. size & Bsize & TPS & Network \\ 
                \midrule
                \textbf{Value} & PBFT & 10 & 5 & 1000 & 0.05 MB & 1MB & 100 &$\mu$:20MB/s, $\sigma$:5MB/s \\
                \bottomrule
            \end{tabular}
        }
    \end{minipage}
    \begin{minipage}[c]{\textwidth}
        \begin{subfigure}[b]{\textwidth}
            \centering
            \includegraphics[width=.8\textwidth]{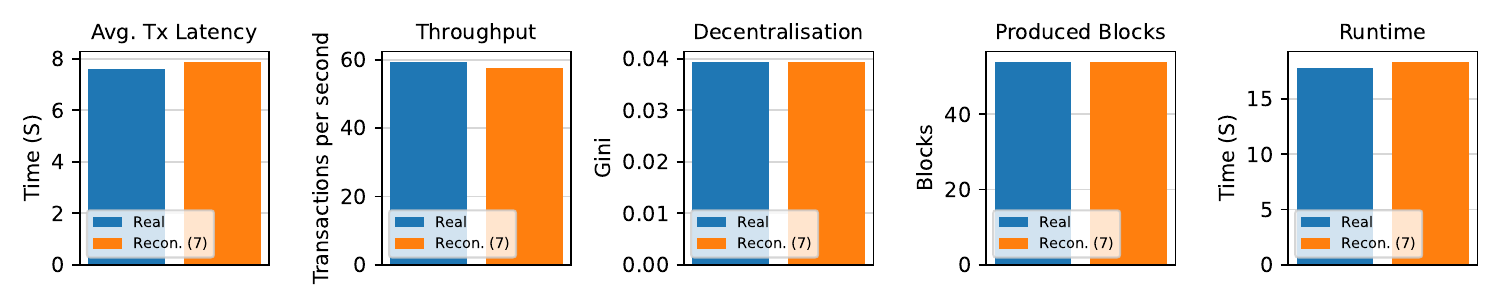} 
            \vspace{-.2cm}
            \caption{Simulation parameters and results (reconstructed global state from 7 state messages)}
            \label{sub:Results}
        \end{subfigure}
    \end{minipage}

    \caption{Simulation results comparing the performance of the real system to that of the reconstructed model}
    \label{fig:results_viz}
\end{figure*}


For the experimental evaluation of the proposed blockchain state extraction algorithm, the blockchain simulation tool SymbChainSim (SBS) \cite{10.1145/3573900.3591121} was utilised to model a permissioned blockchain network running the Practical Byzantine Fault Tolerance (PBFT) consensus protocol. The network architecture of the simulated blockchain was derived by assigning random peers and random bandwidths (modelled as a Gaussian distribution) to the nodes. The number of random peers assigned to each node was strictly less than the total number of nodes in the network. Message propagation took place using SBS's gossip network model, and the workload ($W$) was generated using the simulator's built-in workload generator.

The described system was used to model a physical blockchain to evaluate the ability of the proposed peer-state extraction algorithm. Specifically, the proposed extraction mechanism was utilised to extract the system's state after processing $W$. During the initial simulation, each node monitored all messages exchanged with its peers and constructed the estimated peer state using Eq. \ref{eq:const_eps}. Finally, each node produced a state message containing its local and estimated peer states.

Based on the produced state messages, a set of secondary simulations can be set up to evaluate the proposed algorithm's ability to reconstruct the state of a blockchain. Specifically, these simulations are initialised using a model reconstructed from a subset of the state messages following the proposed reconstruction algorithm. This process models the loss or delay of messages sent over an asynchronous network such as the one in blockchain systems.

\vspace{-.1cm}
\subsection{Results}

Following the setup described above, a blockchain system with $10$ nodes, each with approximately $5$ peers, was initialised and tasked with processing a workload of $1000$ transactions after the simulation was stopped and the state messages of the nodes were extracted. To demonstrate the reconstruction process, 3 state messages were ignored, and their state was reconstructed using the proposed peer-state reconstruction algorithm. The resulting blockchain architecture, the simulation parameters and the simulation results of the original and reconstructed system are shown in Fig \ref{fig:results_viz}.

Focusing on the visualisation of the reconstructed blockchain network architecture, Fig. \ref{sub:Viz}, some interesting observations can be made. Firstly, the algorithm successfully estimated the state of the nodes that failed to provide a state message utilising the information received by their peers. Focusing on node $10$, its bandwidth ($\mu: 24.5, \sigma:  1.92$) was estimated as ($\mu: 21.2, \sigma: 0.35$). The above discrepancy can be attributed to its bandwidth being the highest amongst its peers. Thus, the largest bandwidth of its peers (node 9) acted as the upper bound for that estimation. The same is true for node 6, although to a lesser extent. Node 8 had the best approximation since peers with a higher bandwidth sent their approximation to the digital twin (5,7 and 9). Another important observation is that of the missing edges. The proposed algorithm attempts to reconstruct edges from the peer lists of received state messages. While this can reconstruct edges between nodes that sent a state and nodes that did not (e.g., $8\leftrightarrow7$), edges between nodes that both fail to provide a message cannot be reconstructed. 

The above limitation can be resolved by extending the messages exchanged over the gossip protocol to include information about the entire path the message has taken over the network. This way, a node can reconstruct the edges of nodes further than 1-hop away. Although this technique can reconstruct missing edges, approximating the bandwidth of further neighbours through the message history remains challenging due to the lack of common clocks. Despite the missing edges, the simulation results show that the performance of the reconstructed blockchain system only demonstrates a small discrepancy compared to the original simulation (Fig. \ref{sub:Results}).


\begin{figure*}
\begin{minipage}[c]{1\textwidth}
    \centering
    \resizebox{.55\textwidth}{!}{%
        \begin{tabular}{|l|c|c|c|c|c|c|c|c|}
            \toprule
            \textbf{Param} & $CP$ & Nodes &  Peers per node & Total Tx. & Tx. size & Bsize & TPS & Network\\ 
            \midrule
            \textbf{Value} & PBFT & 50 & 10 & 1000 & 0.05 MB & 1MB & 50 & $\mu$:20MB/s, $\sigma$:5MB/s\\
            \bottomrule
        \end{tabular}
    }
\end{minipage}
\begin{minipage}[c]{\textwidth}
    \begin{subfigure}[b]{\textwidth}
        \includegraphics[width=1\textwidth]{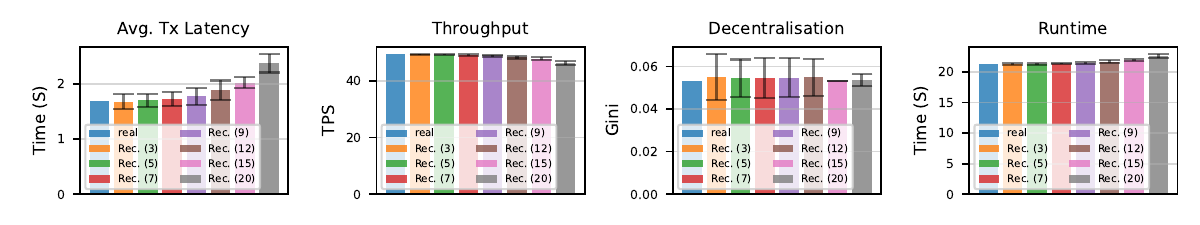} 
    \end{subfigure}
\end{minipage}
\vspace{-.1cm}
\caption{Experimental evaluation of the variance of the simulation results based on the number and specific state messages lost (based on 100 random reconstructions).}
\label{fig:repeats}
\vspace{-.3cm}
\end{figure*}
    
While the results demonstrated in Fig. \ref{fig:results_viz} provided intuition into the proposed algorithm, they also showcased that the quality of the reconstructed state depends on which state messages are lost. For example, if, instead of node 8, the state of node 1 was lost, the algorithm would reconstruct all links in the original system. To evaluate this effect, a blockchain system with 50 nodes was instantiated and used as the physical system from which the state messages for each node were generated. Using the resulting state messages, various models were reconstructed and simulated. Specifically, for 3,5,7,9,12,15 and 20 missing states ($M$), 100 simulations were instantiated by randomly removing $M$ state messages and reconstructing the state using the remaining $N-M$. The results of the above experiment can be seen in Fig. \ref{fig:repeats}. 

The results show that as the number of missing state messages increases, the simulation results diverge more and more from reality. This is natural since the higher the number of missing states, the more information must be reconstructed, which leads to a less accurate model. Another interesting observation is that as the number of missing states increases, the performance of the reconstructed system worsens. This can be attributed to the fact that the upper limit of bandwidth estimations is capped, and thus, there are cases where the estimated bandwidth is lower. The estimation algorithm can also fail to reconstruct missing links (both characteristics are showcased by the reconstruction in Fig. \ref{sub:Viz}). Consequently, missing information tends to result in a slower reconstructed network architecture.

Studying the variance between repeated experiments with random missing states, the observed divination is relatively small and remains consistent as the number of missing states increases. From the above, we can conclude that the accuracy of the reconstructed model mainly depends on the number of missing states. From the evaluated metrics, decentralisation is the least dependent on the network architecture, and as a result, it is not significantly affected by the inaccuracies of the reconstructed network.

\section{Conclusion}
\label{Conclusion}

This paper has considered the challenge of extracting and mirroring the state of a blockchain system to allow for the creation of its Digital Twin. Addressing the problems posed by blockchain's decentralised and asynchronous nature, the paper has proposed a mechanism to reconstruct the state of a blockchain when state updates are lost or delayed. A mechanism to address synchronisation issues has also been proposed. The experimental evaluation has confirmed the robustness of the proposed approach. Future work will consider the presence of malicious nodes in the system and synchronisation issues arising from a distributed simulation model in the  Digital Twin. Finally, studying the upper bounds of the approximation in relation to the number of missing states is

\section*{Acknowledgment}
This research was supported by SUSTech-University of Birmingham Collaborative PhD Programme; Research Institute of Trustworthy Autonomous Systems, Southern University of Science and Technology, Shenzhen 518055, China; Shenzhen Science and Technology Program, the project No.  GJHZ20210705141807022; ZTE Communication Technology Service Co., Ltd., China (Project No: IA20230803007); Georgios Theodoropoulos is the corresponding author.

\bibliographystyle{IEEEtran}
\bibliography{ref.bib}

\end{document}